\documentclass[[12pt,a4paper,preprint,preprintnumbers,amsmath,amssymb,nofootinbib]{revtex4}
\usepackage{graphicx}
\usepackage{amsmath}
\usepackage{amssymb}
\usepackage{bbold}
\usepackage{type1cm}
\usepackage{rotating}
\usepackage{tabularx}
\usepackage{pdflscape}

\usepackage{dsfont}
\usepackage{overpic}
\usepackage{natbib}
\usepackage{slashed}
\usepackage{subfigure}
\usepackage{bbm}

\begin{document}
\title{A toy model for  ``elementariness''}

\author{Peter C.~Bruns}
\affiliation{Nuclear Physics Institute, 25068 \v{R}e\v{z}, Czech Republic }
\begin{abstract}
Motivated by recent efforts to analyze corrections to Weinberg's relations for the scattering length and effective range in the presence of a near-threshold bound state, we play around with an instructive toy model for non-relativistic scattering in a central potential. The model allows to interpolate between bound-state configurations of high ``compositeness'', where the wave function is spread over a wide region beyond the range of the interaction, and compact configurations of high ``elementariness'', where the wave function is confined to a small region around the center of the potential.
\end{abstract}

\maketitle

\section{Introduction}
\label{sec:Intro}

There has been some renewed interest recently \cite{Li:2021cue,Kinugawa:2021ykv,Song:2022yvz,Albaladejo:2022sux} in refinements of the concepts of ``compo\-siteness'' and ``elementariness'' of bound states and resonances \cite{Hyodo:2008xr,Gamermann:2009uq,Hyodo:2011qc,Aceti:2012dd,Hyodo:2013nka,Nagahiro:2014mba,Guo:2015daa,Sekihara:2014kya,Sekihara:2016xnq,Oller:2017alp,Bruns:2019xgo,Baru:2003qq,Baru:2010ww,Guo:2017jvc,Matuschek:2020gqe,Sazdjian:2022kaf}, which can be extracted from scattering data, e.g. \hspace{-0.1cm}through the Weinberg relations \cite{Weinberg:1965zz} for the scattering length and effective range in the case of near-threshold bound states.  In particular, the impact of a non-zero range of the interaction has been under discussion. Having this in mind, it might be amusing, and possibly also instructive, to study these problems in a simple framework familiar from a first course in quantum-mechanical scattering theory, namely, the scattering of non-relativistic particles in a fixed potential $V(r)$ of a finite range (see also \cite{Bruns:2019xgo} for previous work in this direction). \\ Before we specify a model potential, let us shortly recapitulate the formalism, in particular the extraction of ``compositeness'' from partial-wave scattering amplitudes and the Weinberg relations, adapted to our intended model setting.\\ 

We consider elastic s-wave scattering of spinless point particles of mass $\mu$ and energy $E$ in a local, spherically symmetric, energy-independent potential $V(\vec{x})=V(r)$ of finite range. In this case, we can write the s-wave scattering amplitude in the form \cite{Taylor}
\begin{equation}\label{eq:f0gen}
f_{0}(E) = \left[\left(K_{0}(E)\right)^{-1} - ik(E)\right]^{-1} = \left[\left(K_{0}(E)\right)^{-1} + \kappa(E)\right]^{-1}\,.
\end{equation}
Here we employ the notation $k(E)=+\sqrt{2\mu E}=:k$, $\kappa(E)=-ik(E)=:\kappa$. For a bound state $B$ at $E=E_{B}<0$, $k_{B}=i\kappa_{B}=i\sqrt{-2\mu E_{B}}\,$. $K_{0}(E)$ is a meromorphic function of the energy, and real for real $E$. It is well-known from the beginning of S-matrix theory that the residues of scattering amplitudes at the poles pertaining to bound states are related to normalization factors of the bound-state wave functions \cite{Heisenberg:1946ytd}. In our notation,
\begin{equation}\label{eq:f0Res}
f_{0}(E) \,\longrightarrow\, \frac{\mathrm{Res}_{B}f_{0}}{E-E_{B}} = -\frac{\mathcal{N}_{B}^2}{2\mu(E-E_{B})}\quad\mathrm{for}\quad E\rightarrow E_{B}\,,
\end{equation}
when the wave function outside of the range of the potential assumes its asymptotic form $\mathcal{N}_{B}e^{-\kappa_{B}r}\mathcal{Y}_{00}(\theta_{x},\varphi_{x})/r\,$ (of course, $\mathcal{Y}_{00}(\theta_{x},\varphi_{x})=1/\sqrt{4\pi}$, and we use units where $\hbar=c=1$). See e.g. \cite{Bruns:2019xgo} for a demonstration of this relation. \\
If the bound state is located very close to the threshold $E=0$, and $K_{0}^{-1}$ has no poles in the region of such small energies $E\sim E_{B}$, we can match two linear approximations of $\left(K_{0}(E)\right)^{-1}$, namely a truncated Taylor expansion around $E=E_{B}$ and the first two terms in the effective range expansion,
\begin{equation}\label{eq:K0invapprox}
\left(K_{0}(E_{B})\right)^{-1} + (E-E_{B})\frac{dK_{0}^{-1}}{dE}\biggr|_{E_{B}} \overset{!}{\approx} -\frac{1}{a_{0}} + r_{0}\mu E \,,
\end{equation}
which (using $\left(K_{0}(E_{B})\right)^{-1}+\kappa_{B}=0$ and Eq.~(\ref{eq:f0Res})) leads to
\begin{equation}
\kappa_{B}a_{0} \approx \frac{2\mathcal{C}_{B}^{0}}{1+\mathcal{C}_{B}^{0}}\,,\quad \kappa_{B}r_{0} \approx \frac{\mathcal{C}_{B}^{0}-1}{\mathcal{C}_{B}^{0}}\,,
\end{equation}
where the ``compositeness'' $\mathcal{C}_{B}^{0}$ is given by
\begin{equation}\label{eq:defC}
\mathcal{C}_{B}^{0} := -\frac{\mu}{\kappa_{B}}\mathrm{Res}_{B}f_{0} = \frac{\frac{d\kappa}{dE}\bigr|_{E_{B}}}{\frac{dK_{0}^{-1}}{dE}\bigr|_{E_{B}} + \frac{d\kappa}{dE}\bigr|_{E_{B}}}\,,
\end{equation}
or $\,\mathcal{C}_{B}^{0}=\mathcal{N}_{B}^2/(2\kappa_{B})\,$, while the ``elementariness'' $\mathcal{E}_{B}^{0}$ is
\begin{equation}\label{eq:defE}
\mathcal{E}_{B}^{0} := 1-\mathcal{C}_{B}^{0} = \frac{\frac{dK_{0}^{-1}}{dE}\bigr|_{E_{B}}}{\frac{dK_{0}^{-1}}{dE}\bigr|_{E_{B}} + \frac{d\kappa}{dE}\bigr|_{E_{B}}}\,.
\end{equation}
Generalizations of Eqs.~(\ref{eq:defC}) and (\ref{eq:defE}) are extensively discussed in the literature \cite{Li:2021cue,Kinugawa:2021ykv,Song:2022yvz,Albaladejo:2022sux,Hyodo:2008xr,Gamermann:2009uq,Hyodo:2011qc,Aceti:2012dd,Hyodo:2013nka,Nagahiro:2014mba,Guo:2015daa,Sekihara:2014kya,Sekihara:2016xnq,Oller:2017alp}, but for the simple setting studied here, the above relations are adequate. To assess the validity of the linear approximations in Eq.~(\ref{eq:K0invapprox}), we also introduce
\begin{equation}\label{Xar}
  X_{a} := \frac{\kappa_{B}a_{0}}{2-\kappa_{B}a_{0}}\,,\quad X_{r} := \frac{1}{1-\kappa_{B}r_{0}}\,.
\end{equation}
For a sufficiently small $E_{B}$, we should expect $X_{a}\approx X_{r}\approx \mathcal{C}_{B}^{0}$. Of course, it will depend on the details of the potential what ``sufficiently small'' exactly means here.

\section{Toy model potential}

The toy model we shall examine consists of a ``spherical well'' of radius $d$, surrounded by a ``spherical wall'' of thickness $\delta$, i.e.
\begin{equation}
V(r) = V_{0}\theta(d-r) + W_{0}\theta(r-d)\theta(d+\delta-r)\,,\quad V_{0}<0\,,\quad W_{0}\geq 0\,,\quad d>0\,,\quad\delta>0\,,
\end{equation}
where $\theta(\cdot)$ are Heaviside step functions, so the potential vanishes for $r>d+\delta$. It is straightforward to find s-wave bound-state solutions $\psi_{B00}(\vec{x}) = \frac{u_{B00}(r)}{r}\mathcal{Y}_{00}(\theta_{x},\varphi_{x})$ for the Schr\"odinger equation with this potential:
\begin{eqnarray*}
  u_{B00}(r\leq d) &=& \mathcal{N}_{B}e^{-\kappa_{B}(d+\delta)}\frac{\sin(\xi_{B}r)}{\sin(\xi_{B}d)}\left(\left(\frac{\omega+\kappa_{B}}{2\omega}\right)e^{\omega\delta} + \left(\frac{\omega-\kappa_{B}}{2\omega}\right)e^{-\omega\delta}\right)\,,\\
  u_{B00}(d\leq r\leq d+\delta) &=& \mathcal{N}_{B}\left(\left(\frac{\omega+\kappa_{B}}{2\omega}\right)e^{(\omega-\kappa_{B})(d+\delta)}e^{-\omega r} + \left(\frac{\omega-\kappa_{B}}{2\omega}\right)e^{-(\omega+\kappa_{B})(d+\delta)}e^{+\omega r}\right)\,, \\
  u_{B00}(r\geq d+\delta) &=& \mathcal{N}_{B}e^{-\kappa_{B}r}\,,\quad \omega:=\sqrt{2\mu W_{0}+\kappa_{B}^2}\,, \quad \xi_{B} := \sqrt{-2\mu V_{0} - \kappa_{B}^2}\,.
\end{eqnarray*}
The normalization factor $\mathcal{N}_{B}$ is determined  by $\int\,d^3x\,|\psi_{B00}(\vec{x})|^2\overset{!}{=}1$ (as usual, the phase is chosen so that it is real). The bound-state energies must obey the condition
\begin{equation}
\frac{\omega}{\xi_{B}}\tan(\xi_{B}d) + \frac{(\omega+\kappa_{B})e^{\omega\delta} + (\omega-\kappa_{B})e^{-\omega\delta}}{(\omega+\kappa_{B})e^{\omega\delta} - (\omega-\kappa_{B})e^{-\omega\delta}} \overset{!}{=} 0\,,
\end{equation}
and can be found numerically. For the s-wave amplitude $f_{0} = \left[K_{0}^{-1} - ik\right]^{-1}$, one obtains
\begin{equation}
  K_{0} = \frac{1}{k}\,\frac{g_{1}\left(\xi_{V} + \xi_{W}\tan(\xi_{V}d)\tan(\xi_{W}d)\right) + g_{2}\left(\xi_{W}\tan(\xi_{V}d)-\xi_{V}\tan(\xi_{W}d)\right)}{h_{1}\left(\xi_{V} + \xi_{W}\tan(\xi_{V}d)\tan(\xi_{W}d)\right) + h_{2}\left(\xi_{W}\tan(\xi_{V}d)-\xi_{V}\tan(\xi_{W}d)\right)}\,,
\end{equation}
\begin{eqnarray*}
g_{1} &=& k\sin(\xi_{W}(d+\delta))\cos(k(d+\delta))-\xi_{W}\cos(\xi_{W}(d+\delta))\sin(k(d+\delta))\,,\\
g_{2} &=& k\cos(\xi_{W}(d+\delta))\cos(k(d+\delta))+\xi_{W}\sin(\xi_{W}(d+\delta))\sin(k(d+\delta))\,,\\
h_{1} &=& k\sin(\xi_{W}(d+\delta))\sin(k(d+\delta))+\xi_{W}\cos(\xi_{W}(d+\delta))\cos(k(d+\delta))\,,\\
h_{2} &=& k\cos(\xi_{W}(d+\delta))\sin(k(d+\delta))-\xi_{W}\sin(\xi_{W}(d+\delta))\cos(k(d+\delta))\,,
\end{eqnarray*}
where $\xi_{V}:=\sqrt{k^2-2\mu V_{0}}$\,, $\xi_{W}:=\sqrt{k^2-2\mu W_{0}}\,$ ($\xi_{V}\rightarrow \xi_{B}$, $\xi_{W}\rightarrow i\omega$ for a bound state). 
The following table should give an impression of the properties of near-threshold bound states for a chosen parameter set $(V_{0},W_{0},\mu,d,\delta)$.
\begin{center}
  \begin{tabular}{|c|c|c|c|c|c|c|c|}
  \hline
  $W_{0}$ & $V_{0}$ & $\kappa_{B}$ & $\mathcal{C}^{0}_{B}$ & $X_{a}$ &  $X_{r}$ & $\,2\kappa_{B}\langle r\rangle_{B}$ & $P(r>d+\delta)$ \\
  \hline
  0 \quad & \quad -1.284 \qquad & \quad 0.050 \quad & \quad 1.051 \quad & \quad 1.051 \quad & \quad 1.051 \quad  & \quad 1.050 \quad &  0.933 \quad  \\
  3 \quad & \quad -2.159 \qquad & \quad 0.051 \quad & \quad 1.004 \quad & \quad 1.004 \quad & \quad 1.004 \quad  & \quad 1.006 \quad &  0.889 \quad  \\
  15\quad & \quad -3.397 \qquad & \quad 0.047 \quad & \quad 0.692 \quad & \quad 0.695 \quad & \quad 0.698 \quad  & \quad 0.710 \quad &  0.619 \quad  \\
  50\quad & \quad -4.074 \qquad & \quad 0.057 \quad & \quad 0.094 \quad & \quad 0.121 \quad & \quad 0.154 \quad  & \quad 0.151 \quad &  0.082 \quad  \\
  \hline
\end{tabular}
\begin{figure}[h!]
\centering
\includegraphics[width=0.5\textwidth]{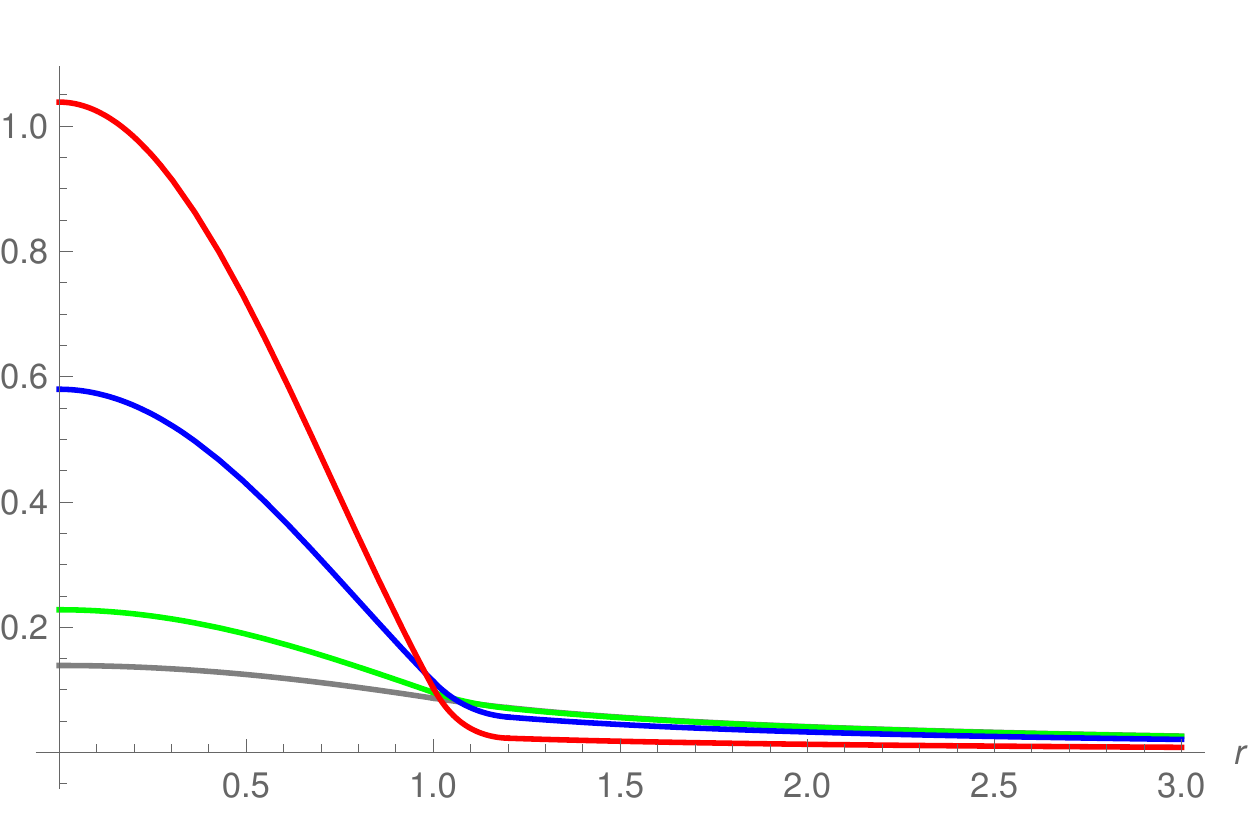}
\caption{The wave functions $\psi_{B00}$ pertaining to the states from the previous table, versus $|\vec{x}|=r$. Gray: $W_{0}=0$, green: $W_{0}=3$, blue: $W_{0}=15$, red: $W_{0}=50$.}
\label{fig:wfs_wellwithwall}
\end{figure}
\end{center}
The examples in the table are for $(\mu,d,\delta)=(1,1,0.2)$. For a chosen $W_{0}$, we have adjusted $V_{0}$ so that a bound state with $\kappa_{B}\approx 0.05\mu$ emerges. $P(r>d+\delta)$ is the probability to find the particle outside of the range of the potential, and $\langle r\rangle_{B}:=\int_{0}^{\infty}dr\,r\,|u_{B00}(r)|^2\,$.\\
The ``elementariness'' $1-\mathcal{C}^{0}_{B}$ will approach one when the ``wall height'' $W_{0}\rightarrow +\infty\,$. Then, 
\begin{displaymath}
u_{B00}(r)\rightarrow\sqrt{\frac{2}{d}}\,\theta(d-r)\sin\left(\frac{\pi r}{d}\right)\,,\quad \langle r\rangle_{B}\rightarrow\frac{d}{2}\,,
\end{displaymath}
the bound particle becomes confined to the region $r<d$ and is shielded from the outside world, while the s-wave amplitude approaches that for scattering on a hard sphere. This appearance of a small impermeable shell is the closest we can get to ``pure elementariness'' in a potential model.\\
To make sure that these observations do not depend crucially on peculiar features of the potential like e.g. the discontinuities at $r=d$ and $r=d+\delta$, we consider in App.~\ref{app:osc3Dwellwall} a continuous potential of a similar ``well+wall'' structure, assembled from harmonic-oscillator potentials. It is seen that the properties of the bound states qualitatively agree with those of the model studied above.

\section{Conclusions}
\label{sec:conclusions}
While it might at first seem counterintuitive that one can have a noticeable ``elementariness'' in a potential-scattering model, it is easily explained.
Due to the relation \cite{Bruns:2019xgo} $\mathcal{C}_{B}^{0}=e^{2\kappa_{B}R}P(r>R)$ for the ``compositeness'' (as defined in Eq.~(\ref{eq:defC})), where $R$ is the finite range of the potential (in our case, $R=d+\delta$), it is clear that one can find states of arbitrarily small $\mathcal{C}_{B}^{0}$ in a potential that prevents the wave function of the low-energy bound state from ``leaking out'' to the outside region where $V(r)=0$. The toy model presented in this contribution is a simple example where this can be accomplished. Put differently, the model potential simulates the presence of a real elementary state. The challenge would be to distinguish one possibility from the other for a given set of scattering data. It is conceivable that the model discussed here could serve as a good testing ground for the extraction and interpretation of ``elementariness'' from scattering amplitudes, before one goes on to deal with more realistic, and much more difficult, problems involving relativistic dynamics, energy-dependent potentials, inelasticity, resonances and so on. For these more subtle questions we refer to the cited literature.

%

\newpage

\begin{appendix}

\section{Harmonic well with a harmonic wall}
\label{app:osc3Dwellwall}
\def\theequation{\Alph{section}.\arabic{equation}}
\setcounter{equation}{0}

Potential: $V(r) = \theta(d-r)V_{0}\left(1-\frac{r^2}{d^2}\right) + \theta(r-d)\theta(d+\delta-r)W_{0}\left(1-\frac{4}{\delta^2}(r-(d+(\delta/2)))^2\right)\,$,\\
with $V_{0}<0$, $W_{0}\geq 0$. We employ the abbreviations $\,\kappa_{B}=+\sqrt{-2\mu E_{B}}\,$ and
\begin{equation*}
  \zeta_{V} = \sqrt{-2\mu V_{0} -\kappa_{B}^2}\,,\quad \zeta_{W} = \sqrt{-2\mu W_{0} -\kappa_{B}^2}\,,\quad
  \lambda_{V} = \frac{\sqrt{-2\mu V_{0}}}{d}\,,\quad  \lambda_{W} = \frac{2\sqrt{-2\mu W_{0}}}{\delta}\,.
\end{equation*}
\vspace{-0.55cm}
\begin{figure}[h!]
\centering
\includegraphics[width=0.48\textwidth]{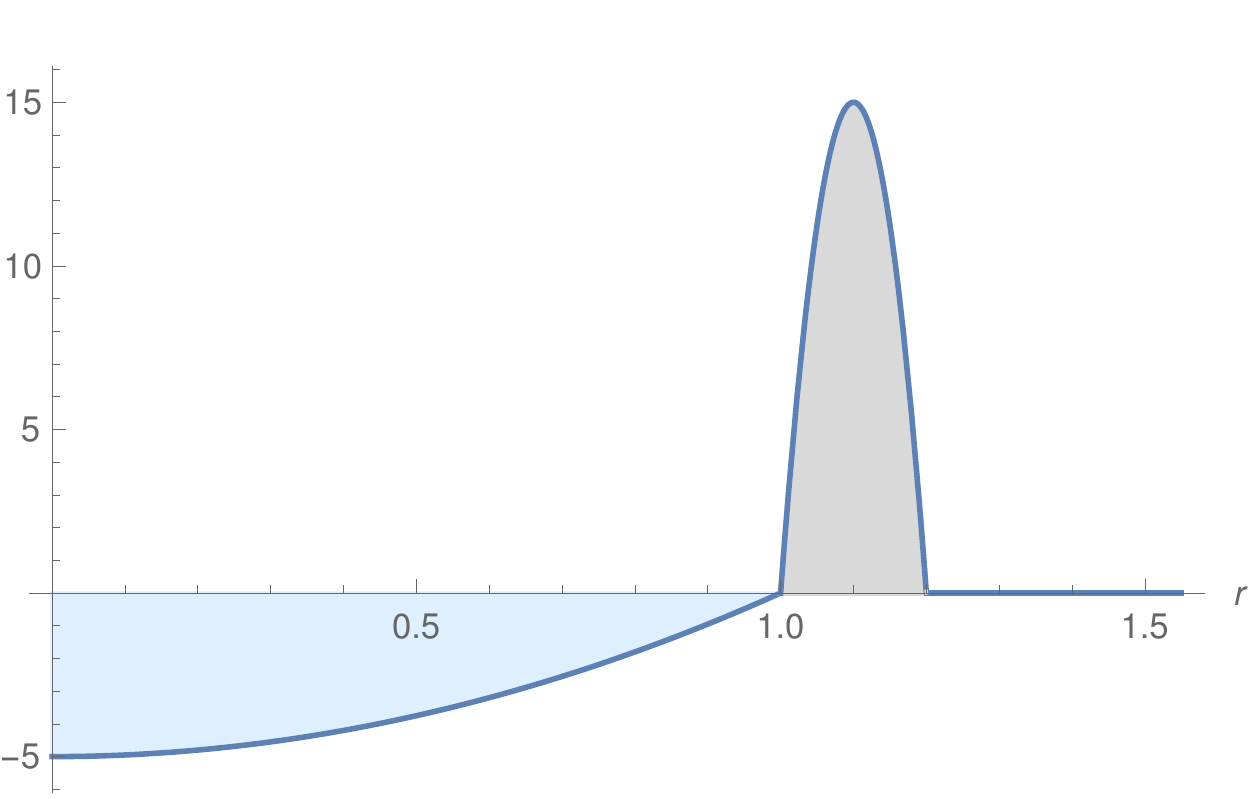}
\caption{The potential $V(r)$ for $V_{0}=-5$, $W_{0}=15$, $d=1$, $\delta=0.2$ (in units of $\mu=1$).}
\label{fig:oscwellwithwall}
\end{figure}\\
\quad
The s-wave solutions can be found as $\psi_{B00}(\vec{x}) = \frac{u_{B00}(r)}{r}\mathcal{Y}_{00}(\theta_{x},\varphi_{x})$, with
\begin{eqnarray*}
  u_{B00}(r\leq d) &=& C_{B}re^{-\frac{1}{2}\lambda_{V}r^2}\,_1\hspace{-0.05cm}F_{1}\left(\frac{1}{4}\left(3-\frac{\zeta_{V}^2}{\lambda_{V}}\right),\,\frac{3}{2},\,\lambda_{V} r^2\right)\,,\\
  u_{B00}(d\leq r\leq d+\delta) &=& C_{1}\tilde{r}e^{-\frac{1}{2}\lambda_{W}\tilde{r}^2}\,_1\hspace{-0.05cm}F_{1}\left(\frac{1}{4}\left(3-\frac{\zeta_{W}^2}{\lambda_{W}}\right),\,\frac{3}{2},\,\lambda_{W} \tilde{r}^2\right)\\
  &+& C_{2}\,e^{-\frac{1}{2}\lambda_{W}\tilde{r}^2}\,_1\hspace{-0.05cm}F_{1}\left(\frac{1}{4}\left(1-\frac{\zeta_{W}^2}{\lambda_{W}}\right),\,\frac{1}{2},\,\lambda_{W} \tilde{r}^2\right)\,, \quad \tilde{r}:=r-\left(d+\frac{\delta}{2}\right)\,, \\
  u_{B00}(r\geq d+\delta) &=& \mathcal{N}_{B}e^{-\kappa_{B}r}\,.
\end{eqnarray*}
As usual, the constants $C_{1,2}$ and $C_{B}$ as well as possible values of $\kappa_{B}$ are found via the constraints of continuity and differentiability at $r=d$ and $r=d+\delta$; $\mathcal{N}_{B}$ is subsequently fixed by the normalization condition. The following table displays some properties of the least-bound state, for different ``wall heights'' $W_{0}=0,3,15$ and $50$, and $d=1$, $\delta=0.2\,$.\\
\begin{center}
  \vspace{-0.25cm}
  \begin{tabular}{|c|c|c|c|c|c|c|c|}
  \hline
  $W_{0}$ & $V_{0}$ & $\kappa_{B}$ & $\mathcal{C}^{0}_{B}$ & $X_{a}$ & $X_{r}$ & $\,2\kappa_{B}\langle r\rangle_{B}$ & $P(r>d+\delta)$ \\
  \hline
  0 \quad & \quad -2.647 \qquad &  \quad 0.050 \quad & \quad 1.042 \quad & \quad 1.042 \quad & \quad 1.042 \quad & \quad 1.041 \quad  & 0.924  \\
  3 \quad & \quad -3.613 \qquad &  \quad 0.050 \quad & \quad 1.000 \quad & \quad 1.000 \quad & \quad 1.000 \quad & \quad 1.002 \quad  & 0.887  \\
  15\quad & \quad -4.984 \qquad &  \quad 0.049 \quad & \quad 0.752 \quad & \quad 0.754 \quad & \quad 0.756 \quad & \quad 0.767 \quad  & 0.668  \\
  50\quad & \quad -5.740 \qquad &  \quad 0.048 \quad & \quad 0.206 \quad & \quad 0.223 \quad & \quad 0.242 \quad & \quad 0.247 \quad  & 0.184  \\
  \hline
\end{tabular}
\end{center}

\newpage

\section{The case of the deuteron}
\label{app:deuteron}
\def\theequation{\Alph{section}.\arabic{equation}}
\setcounter{equation}{0}

Just for the time being, we shall assume that the deuteron can be described with a finite-range potential of the kind considered in this work. For the deuteron, we have
\begin{equation}\label{eq:deuteron_data}
\mu = \frac{m_{n}m_{p}}{m_{n}+m_{p}} \approx 0.47\,\mathrm{GeV}\,, \quad \kappa_{B} \approx \frac{\mu}{10.272}\,,\quad X_{a}\approx 1.68\,,\quad X_{r}\approx 1.69\,.
\end{equation}
We can already say that our treatment will at least be  more realistic than with a zero-range interaction, because from our formula for $\mathcal{C}_{B}^{0}$ (see Sec.~\ref{sec:conclusions}) we can derive the following estimate for the lower bound of the range,
\begin{equation}
R \gtrapprox \frac{\ln\left(1.68\right)}{2\kappa_{B}} \approx 2.66\,\mu^{-1} \approx 0.79\,M_{\pi}^{-1}\,,
\end{equation}
since $P(r>R)\leq 1\,$. Compare also \cite{Song:2022yvz} for a very recent study underlining the importance of the interaction range for the deuteron case. More concretely, let us adopt the harmonic potential of App.~\ref{app:osc3Dwellwall} with $W_{0}=0$, so that effectively $R=d$. One then finds that one can reproduce the data of Eq.~(\ref{eq:deuteron_data}) with
\begin{equation}
V_{0} = -0.098\,\mu\,,\, d = 6.168\,\mu^{-1} \quad \Rightarrow \quad \mathcal{C}_{B}^{0} = 1.666 \quad \Rightarrow \quad P(r>d) = 0.502\,.
\end{equation}
This would really be ``evidence that the deuteron is not an elementary particle'' \cite{Weinberg:1965zz}. \\
\quad \\
Why could one say this? Here is an (admittedly crude) analogy: Why am I confident that a couple in my neighborhood, named Smetana, is composed of two individuals labeled as ``Mr.~Smetana'' and ``Mrs.~Smetana'', and is not just a single organism one could call e.g.\hspace{-0.2cm} a ``Smetanovi''? Well, in the majority of times I see them, I perceive two objects with the typical characteristics of a human being, and I estimate their distance to be more than two arm lengths,
\begin{displaymath}
  P(r>2\ell(\mathrm{arm}))>0.5\,.
\end{displaymath}
In such an instance, one can in principle examine one of the components individually (put it in a magnetic field, weigh it, turn it upside down, bombard it with antineutrinos, etc.).\\
Sometimes, however, quite rarely, I see them from a distance being so close together that I could suspect that they have merged into one two-headed creature with four legs\,...\\
But most likely I just need a better telescope.

\end{appendix}


\newpage

\begin{thebibliography}{99}

  
\bibitem{Li:2021cue}
Y.~Li, F.~K.~Guo, J.~Y.~Pang and J.~J.~Wu,
[arXiv:2110.02766 [hep-ph]].

\bibitem{Kinugawa:2021ykv}
T.~Kinugawa and T.~Hyodo,
[arXiv:2112.00249 [hep-ph]].

\bibitem{Song:2022yvz}
J.~Song, L.~R.~Dai and E.~Oset,
[arXiv:2201.04414 [hep-ph]].

\bibitem{Albaladejo:2022sux}
M.~Albaladejo and J.~Nieves,
[arXiv:2203.04864 [hep-ph]].


\bibitem{Hyodo:2008xr}
  T.~Hyodo, D.~Jido and A.~Hosaka,
  Phys.\ Rev.\ C {\bf 78} (2008) 025203
  [arXiv:0803.2550 [nucl-th]].
  

\bibitem{Gamermann:2009uq}
  D.~Gamermann, J.~Nieves, E.~Oset and E.~Ruiz Arriola,
  Phys.\ Rev.\ D {\bf 81} (2010) 014029.

\bibitem{Hyodo:2011qc}
T.~Hyodo, D.~Jido and A.~Hosaka,
Phys. Rev. C \textbf{85} (2012), 015201
[arXiv:1108.5524 [nucl-th]].
  
\bibitem{Aceti:2012dd}
  F.~Aceti and E.~Oset,
  Phys.\ Rev.\ D {\bf 86} (2012) 014012
  [arXiv:1202.4607 [hep-ph]].

  
\bibitem{Hyodo:2013nka}
  T.~Hyodo,
  Int.\ J.\ Mod.\ Phys.\ A {\bf 28} (2013) 1330045
  [arXiv:1310.1176 [hep-ph]].
  
\bibitem{Nagahiro:2014mba}
  H.~Nagahiro and A.~Hosaka,
  Phys.\ Rev.\ C {\bf 90} (2014) no.6,  065201
  [arXiv:1406.3684 [hep-ph]].

\bibitem{Guo:2015daa}
  Z.~H.~Guo and J.~A.~Oller,
  Phys.\ Rev.\ D {\bf 93} (2016) no.9,  096001
  [arXiv:1508.06400 [hep-ph]].


\bibitem{Sekihara:2014kya}
  T.~Sekihara, T.~Hyodo and D.~Jido,
  PTEP {\bf 2015} (2015) 063D04
  [arXiv:1411.2308 [hep-ph]].
  
  
\bibitem{Sekihara:2016xnq}
  T.~Sekihara,
  Phys.\ Rev.\ C {\bf 95} (2017) no.2,  025206
  [arXiv:1609.09496 [quant-ph]].

\bibitem{Oller:2017alp}
  J.~A.~Oller,
  Annals Phys.\  {\bf 396} (2018) 429
  [arXiv:1710.00991 [hep-ph]].

\bibitem{Bruns:2019xgo}
P.~C.~Bruns,
[arXiv:1905.09196 [hep-ph]].  
  
  

 

  









\bibitem{Baru:2003qq}
  V.~Baru, J.~Haidenbauer, C.~Hanhart, Y.~Kalashnikova and A.~E.~Kudryavtsev,
  Phys.\ Lett.\ B {\bf 586} (2004) 53
  [hep-ph/0308129].



\bibitem{Baru:2010ww}
  V.~Baru, C.~Hanhart, Y.~Kalashnikova, A.~E.~Kudryavtsev and A.~V.~Nefediev,
  Eur.\ Phys.\ J.\ A {\bf 44} (2010) 93
  [arXiv:1001.0369 [hep-ph]].

\bibitem{Guo:2017jvc}
  F.~K.~Guo, C.~Hanhart, U.-G.~Mei{\ss}ner, Q.~Wang, Q.~Zhao and B.~S.~Zou,
  Rev.\ Mod.\ Phys.\  {\bf 90} (2018) no.1,  015004
  [arXiv:1705.00141 [hep-ph]].


\bibitem{Matuschek:2020gqe}
I.~Matuschek, V.~Baru, F.~K.~Guo and C.~Hanhart,
Eur. Phys. J. A \textbf{57} (2021) no.3, 101
[arXiv:2007.05329 [hep-ph]].

\bibitem{Sazdjian:2022kaf}
H.~Sazdjian,
Symmetry \textbf{14} (2022), 515 
[arXiv:2202.01081 [hep-ph]].


\bibitem{Weinberg:1965zz}
  S.~Weinberg,
  Phys.\ Rev.\  {\bf 137} (1965) B672.

  
\bibitem{Taylor}
 J.~R.~Taylor,
 ``Scattering Theory'',
 John Wiley \& Sons (1972).

\bibitem{Heisenberg:1946ytd}
W.~Heisenberg,
Z. Naturforsch. \textbf{1} (1946) no.11-12, 608-622.

  




  




 

 

  
 

 



  
\end{thebibliography}
\end{document}